\documentclass[aps,prl,amsmath, amssymb, twocolumn,
nofootinbib,showpacs,tightenlines]{revtex4}
\global\arraycolsep=2pt
\usepackage{bm}

\newcommand{\ben}{\begin{equation*}}
\newcommand{\een}{\end{equation*}}
\newcommand{\bean}{\begin{eqnarray*}}
\newcommand{\eean}{\end{eqnarray*}}

\newcommand{\nn}{\nonumber}
\newcommand{\be}{\begin{equation}}
\newcommand{\ee}{\end{equation}}
\newcommand{\bea}{\begin{eqnarray}}
\newcommand{\eea}{\end{eqnarray}}
\newcommand{\binomial}[2]{ \begin{pmatrix}#1\\#2\end{pmatrix} }
\begin{document}
\title{Exact Casimir Interaction Between Semitransparent Spheres and
Cylinders}

\author{Kimball A. Milton} 
\email{milton@nhn.ou.edu}
\homepage{http://www.nhn.ou.edu/%7Emilton}

\author{Jef Wagner}
\email{wagner@nhn.ou.edu}
\affiliation{Oklahoma Center for High Energy Physics 
and Homer L. Dodge Department of Physics and Astronomy,
University of Oklahoma, Norman, OK 73019, USA}
\date{\today}
\pacs{03.70.+k, 03.65.Nk, 11.80.Et, 11.80.La}

\begin{abstract}
A multiple scattering formulation is used to calculate the force, arising
from fluctuating scalar fields,
between distinct bodies described by $\delta$-function potentials, so-called
semitransparent bodies. (In the limit of strong coupling, a semitransparent
boundary becomes a Dirichlet one.) We obtain expressions for the Casimir
energies between disjoint parallel semitransparent cylinders 
and between disjoint semitransparent spheres. In the limit of weak
coupling, we derive power series expansions for the energy, which can be
exactly summed, so that explicit, very simple, closed-form expressions
are obtained in both cases.  The proximity force theorem holds
when the objects are almost touching, but  is subject to large corrections
as the bodies are moved further apart.
\end{abstract}

\maketitle

Multiple scattering methods for calculating Casimir (quantum vacuum)
energies between bodies date back to the famous papers of Balian and 
Duplantier \cite{Balian:1977qr,Balian:1976za,Balian:2004jv}. More
recently, Emig and collaborators \cite{buscher,Emig:2006uh} 
have published a series of papers, using closely related methods,
to calculate numerically forces between distinct bodies,
starting from periodically deformed ones \cite{Emig:2002xz}.
The methods were developed independently in papers by 
Bulgac, Marierski, and Wirzba \cite{Bulgac:2005ku,Wirzba:2005zn,wirzba07}, 
who, using the Krein formula \cite{krein}, obtained 
results for the interaction of two spheres, 
or a sphere and a plate (for Dirichlet boundary conditions),
and by Bordag \cite{Bordag:2006vc} who rederived the methods
developed in Refs.~\cite{Emig:2006uh,Bulgac:2005ku} in a modified
form and applied them to calculate the first correction beyond the
proximity force approximation (PFA) for a cylinder near a plane.
In Ref.~\cite{Bordag:2006kx} Bordag generalized the method of
Ref.~\cite{Bordag:2006vc} to the case of a semitransparent cylinder
next to a plane. Dalvit et al.\ 
\cite{Mazzitelli:2006ne, Dalvit:2006wy}
used the argument principle to calculate the interaction between
conducting cylinders with parallel axes when one cylinder is inside the other.
Recently, there appeared papers concerning ``exact'' methods of
calculating Casimir energies or forces between arbitrary distinct bodies
by Emig, Graham, Jaffe, and Kardar \cite{Emig:2007cf,Emig:2007me}.
See also Refs.~\cite{Emig:2007qw,sidewalls}. 
Most explicitly, an earlier drafted paper by Kenneth and Klich 
\cite{Kenneth:2007jk} appeared which shows that
 the basis of the latter approach lies in the Lippmann-Schwinger formulation
of scattering theory \cite{lippmann}.  

We might also mention the worldline numerical method of  Gies and Klingm\"uller 
\cite{Gies:2006bt,Gies:2006xe,Gies:2006cq,Gies:2003cv},
but that method lies rather outside our discussion here. 
The same applies to the work of Capasso et al.~\cite{capasso}, who
calculate forces from stress tensors using the familiar construction
of the stress tensor in terms of Green's
dyadics \cite{Schwinger:1977pa,Milton:1978sf}, using a numerical 
engineering method: Finite-difference
frequency-domain methods are employed in two dimensions to obtain forces
between metal squares and plates to 3\% accuracy.

We will now proceed to restate the multiple scattering technique,
in a simple, straightforward way, and apply it to various situations, 
all characterized by $\delta$-function potentials.

The general formula for the Casimir energy (for simplicity here we restrict 
attention to a massless scalar field) is \cite{Schwinger75}
\be
E=\frac{i}{2\tau}\mbox{Tr}\ln G\to 
\frac{i}{2\tau} \mbox{Tr}\ln G G_0^{-1},\label{trln}
\ee
where $\tau$ is the ``infinite'' time that the
configuration exists, and $G$ is the Green's function in the
presence of a potential $V$ satisfying 
(matrix notation)
\be
(-\partial^2+V)G=1, \ee
subject to some boundary conditions at infinity.  (Details will
be supplied elsewhere \cite{scat}.)
In the second form of Eq.~(\ref{trln}) we have subtracted the 
energy of the vacuum, by
inserting the free Green's function $G_0$, which satisfies, 
with the same boundary conditions as $G$, the
free equation
\be -\partial^2 G_0=1.\ee
Now we define the $T$-matrix (note that our definition of $T$ differs by
a factor of 2 from that in Ref.~\cite{Emig:2007cf})
\be
T=S-1=V(1+G_0V)^{-1}.\ee
We then follow standard scattering theory \cite{lippmann}, as reviewed
in Kenneth and Klich \cite{Kenneth:2007jk}. 
If the potential has two disjoint parts, $V=V_1+V_2$,
it is easy to derive the following
general expression for the interaction between
two bodies (potentials):
\be
E_{12}= -\frac{i}{2\tau}\mbox{Tr}\ln(1-G_0T_1G_0T_2),\label{gtgt}
\ee
where
\be
T_i=V_i(1+G_0V_i)^{-1},\quad i=1,2.\ee
This form is exactly that given by Emig et al. \cite{Emig:2007cf},
and by Kenneth and Klich \cite{Kenneth:2007jk}.  In passing from
Eq.~(\ref{trln}) to Eq.~(\ref{gtgt}) we have removed self-action
terms that are generally divergent, but that do not refer to the
separation between the bodies.

Elsewhere \cite{scat} we will show that this formulation allows us to
rederive the Casimir interaction between two semitransparent plates,
and the self-energy of the semitransparent sphere.

\section{$2+1$ Spatial Dimensions} 
We now proceed to apply this method to the interaction between bodies, 
starting with a $2+1$ dimensional version, which allows us to describe, 
for example, cylinders with parallel axes.  Let the distance between the
centers of the bodies be $R$. Then we perform a Fourier
analysis of the reduced Green's function, defined by
\be
G_0(\mathbf{R+r'-r})%=\frac{e^{i|\omega||\mathbf{r-R-r'}|}}{4\pi|
%\mathbf{r-R-r'}|}
=\int\frac{dk_z}{2\pi}e^{ik_z(z-z')}g_0(\mathbf{R_\perp+r'_\perp-r_\perp}),
\ee
where the reduced Green's function has the expansion
(as long as the two potentials do not overlap) 
\be
g_0%(\mathbf{r_\perp-R_\perp-r'_\perp})
=\sum_{m,m'}I_m(\kappa r) e^{im\phi}
I_{m'}(\kappa r')e^{-im'\phi'}
\tilde g^0_{m,m'}(\kappa R),\ee
where $\omega=i\zeta$ and $\kappa^2=k_z^2+\zeta^2$.  The Fourier-Bessel
transform of the reduced Green's function is
\be
\tilde g_{m,m'}^0(\kappa R)=\frac{(-1)^{m'}}{2\pi}K_{m-m'}(\kappa R).
\ee
Thus we can derive from Eq.~(\ref{gtgt}) 
an expression for the interaction energy (per unit length $L$)
between two bodies, in terms of discrete matrices,
\be
\frac{E_{12}}{L}=\frac1{8\pi^2}\int d\zeta\,dk_z\ln\det\left(1-\tilde g^0
t_1\tilde g^{0\top} t_2\right),\ee
where $\top$ denotes transpose, and
where the $t$ matrix elements are given by
\be
t_{mm'}=\int (d\mathbf{r_\perp}) \int (d\mathbf{r'_\perp})I_m(\kappa r)
e^{-im\phi}I_{m'}(\kappa r')e^{im'\phi'}T.\ee

Consider, as an example, two parallel semitransparent cylinders, of
radii $a$ and $b$, respectively, lying
outside each other, described by the potentials
\be
V_1=\lambda_1 \delta(r-a),\quad V_2=\lambda_2\delta(r'-b),
\ee
with the separation $R$ between the axes satisfying $R>a+b$.
It is easy to work out the scattering matrix:
\be
(t_1)_{mm'}=2\pi\lambda_1a\delta_{mm'}\frac{I_m^2(\kappa a)}{1+\lambda_1 a
I_m(\kappa a)K_m(\kappa a)}.\label{met}
\ee
Then the Casimir energy $E$ is
\be
\frac{E}L=\frac1{4\pi}\int_0^\infty d\kappa\,\kappa\,\mbox{tr}\ln(1-A),
\label{eylenergy}
\ee
where $A=B(a)B(b)$,
in terms of the matrices
\be
B_{mm'}(a)=K_{m+m'}(\kappa R)\frac{\lambda_1 a I_{m'}^2(\kappa a)}
{1+\lambda_1 aI_{m'}(\kappa a)K_{m'}(\kappa a)}.\label{bofa}
\ee

As a check, it is easy to reproduce the result derived by 
Bordag \cite{Bordag:2006vc} for a cylinder in front of a plane, using
an evident image method.

\begin{figure}[t]
\vspace{2.in}
\includegraphics{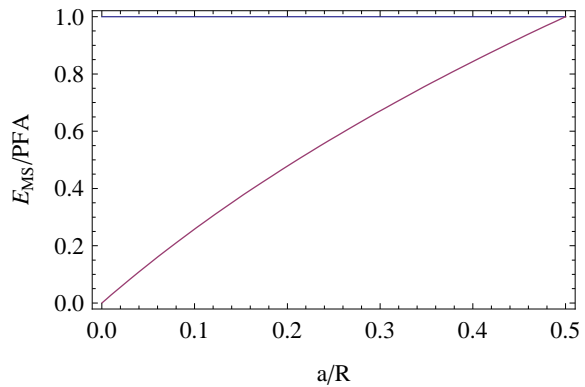}
\caption{\label{figwccyl1}
Plotted is the ratio of the exact interaction energy
(\ref{excyl}) of two weakly-coupled
cylinders to the proximity force approximation 
(\ref{pfawccyl}) as a function of the cylinder radius $a$ 
for $a=b$.}
\end{figure}

In weak coupling, the formula (\ref{eylenergy})
for the interaction energy between the cylinders
is
\bea
\frac{E}L&=&-\frac{\lambda_1\lambda_2 ab}{4\pi R^2}\sum_{m,m'=-\infty}^\infty\nn\\
&&\times\int_0^\infty dx\,x\, K_{m+m'}^2(x)I_m^2(xa/R)I_{m'}^2(xb/R).
\eea
It is straightforward to develop a power series in $a/R$ for the interaction
between semitransparent cylinders.  One merely exploits the small argument
expansion for the modified Bessel functions $I_m(xa/R)$ and $I_{m'}(xb/R)$.
The result is amazingly simple:
\be
\frac{E}L=-\frac{\lambda_1a\lambda_2b}{4\pi R^2}\frac12\sum_{n=0}^\infty
\left(\frac{a}{R}\right)^{2n}P_n\left(\frac{b}{a}\right),\label{multicyl}
\ee
where in terms of the binomial coefficients
\be
P_n\left(\frac{b}{a}\right)=\sum_{k=0}^n \left(\begin{array}{c}n\\k\end{array}
\right)^2\left(\frac{b}a\right)^{2k}.
\ee
Remarkably, it is possible to perform the sums \cite{riordan}, 
so we obtain the following
closed form for the interaction between two weakly-coupled cylinders:
\be
\frac{E}{L}=-\frac{\lambda_1 a \lambda_2 b}{8 \pi R^2}
\left[\!\! \left( 1 - \left( \frac{a+b}{R} \right)^2\right)\!\!
\left( 1 - \left( \frac{a-b}{R} \right)^2 \right)\!\! \right]^{-1/2}.
\label{excyl}
\ee
We note that in the limit $R-a-b=d\to0$, $d$ being the distance between
the closest points on the two cylinders, we recover the proximity force
theorem in this case,
\be V(d)=-\frac{\lambda_1\lambda_2 }{32\pi}\sqrt{\frac{2ab}R}\frac1{d^{1/2}},
\quad d\ll a, b.
\label{pfawccyl}
\ee
In Figs.~\ref{figwccyl1}--\ref{figwccyl2} 
we compare the exact energy (\ref{excyl}) with the
(ambiguously defined) proximity force approximation (\ref{pfawccyl}).  
Evidently, the former approach the latter
when the sum of the radii $a+b$ of the cylinders approaches the distance $R$
between their centers. The rate of approach is linear (with slope 3/2)
for the equal radius
case, but with slope $b^2/4a^2$ when $a\ll b$.

\begin{figure}[t]
\vspace{2in}
\includegraphics{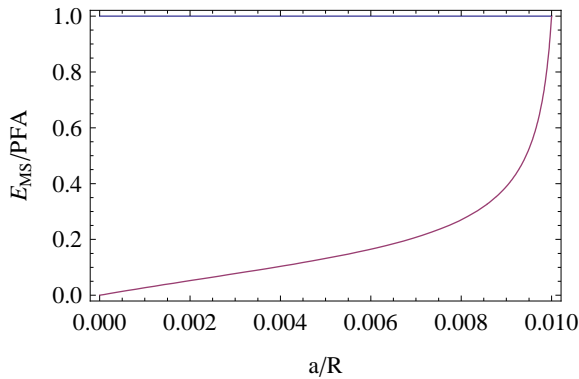}
\caption{\label{figwccyl2}
Plotted is the ratio of the exact interaction energy
(\ref{excyl}) of two weakly-coupled
cylinders to the proximity force approximation 
(\ref{pfawccyl}) as a function of the cylinder radius $a$ 
for $b/a=99$.}
\end{figure}

\section{3-dimensional formalism}

The three-dimensional formalism is very similar.  Again, details will
be supplied in Ref.~\cite{scat}.  Let us proceed to write down the
expression following from Eq.~(\ref{gtgt})
for the interaction between two semitransparent cylinders: 
\be
E=\frac1{4\pi}\int_0^\infty d\zeta \,\mbox{tr}\ln(1-A),\ee
where the matrix
\be
A_{lm,l'm'}=\delta_{m,m'}\sum_{l''}B_{ll''m}(a)B_{l''l'm}(b)
\ee
is given in terms of the quantities [the three-$j$ symbols (Wigner coefficients) 
here vanish unless $l+l'+l''$ is even]
\begin{widetext}
\be
B_{ll'm}(a)=
\frac{i\sqrt{\pi}}{\sqrt{2\zeta R}} i^{-l+l'} \sqrt{(2l+1)(2l'+1)}
\sum_{l''}(2l''+1)
\left(\begin{array}{ccc}l&l'&l''\\0&0&0\end{array}
\right)\left(\begin{array}{ccc}l&l'&l''\\m&-m&0\end{array}\right)
\frac{K_{l''+1/2}(\zeta R)\lambda_1 a I_{l'+1/2}^2(\zeta a)}
{1+\lambda_1 a I_{l'+1/2}(\zeta a)K_{l'+1/2}(\zeta a)}.
\ee
%\end{widetext}
For strong coupling, this result reduces to that found by Bulgac
et al.~\cite{Bulgac:2005ku} for Dirichlet spheres, and recently
generalized by Emig et al.~\cite{Emig:2007me} for Robin boundary conditions.

For weak coupling, a major simplification results because ot the
orthogonality property ($l\le l'$),
\be
\sum_{m=-l}^l\left(\begin{array}{ccc}l&l'&l''\\m&-m&0\end{array}
\right)\left(\begin{array}{ccc}l&l'&l'''\\m&-m&0\end{array}\right)
=\delta_{l''l'''}\frac1{2l''+1}.
\ee
Then the formula for the energy of interaction between the two spheres is
%\begin{widetext}
\be
E=-\frac{\lambda_1a\lambda_2b}{4R}\int_0^\infty \frac{dx}x\sum_{ll'l''}
(2l+1)(2l'+1)(2l''+1)
\left(\begin{array}{ccc}l&l'&l''\\0&0&0\end{array}\right)^2
K_{l''+1/2}^2(x)I_{l+1/2}^2(xa/R)I_{l'+1/2}^2(xb/R).
\ee
\end{widetext}
There is no infrared divergence because for small $x$ the product of Bessel
functions goes like $x^{2(l+l'-l'')+1}$, and $l''\le l+l'$ because of the
triangle property of the 3$j$-symbols. 

Again, it is straightforward to carry out a power series expansion in $a/R$,
which turns out to have a simple form
\bea
E&=&- \frac{\lambda_1 a \lambda_2 b }{8 R}
\frac{a b}{R^2}
\sum_{n=0}^\infty \frac{1}{n+1} \sum_{m=0}^n
\binomial{2n+2}{2m+1}\nn\\
&&\quad\times
\left(\frac{a}{R}\right)^{2n}\left(\frac{b}{a}\right)^{2m}.\label{mesphere}
\eea
Once more, it can be recognized as the following closed form:
\be
E=\frac{\lambda_1 a \lambda_2 b}{16 \pi R} \text{ln}\left(
\frac{1-\left(\frac{a}{R}+\frac{b}{R}\right)^2}{1-\left(\frac{a}{R}-\frac{b}
{R}\right)^2}
\right).\label{exsphere}
\ee
Again, when $d=R-a-b\ll a,b$, the proximity force theorem is reproduced:
\be
V(d)\sim \frac{\lambda_1\lambda_2ab}{16\pi R}\ln (d/R),\quad
d\ll a, b.\label{pfawcsphere}
\ee  
However, as Figs.~\ref{fig3}, \ref{fig4} demonstrate, the
approach is not very smooth,
even for equal-sized spheres. The ratio of the energy to the PFA is
\be
\frac{E}{V}=1+\frac{\ln[(1+\alpha)^2/2\alpha]}{\ln d/R},\quad d\ll a, b,
\ee
for $b/a=\alpha$. Truncating the power series (\ref{mesphere}) at $n=100$
would only begin to show the approach to the PFA limit.  The error in using
the PFA formula between spheres (\ref{pfawcsphere}) outside its range
of validity can be very substantial.

\begin{figure}[t]
\vspace{2in}
\includegraphics{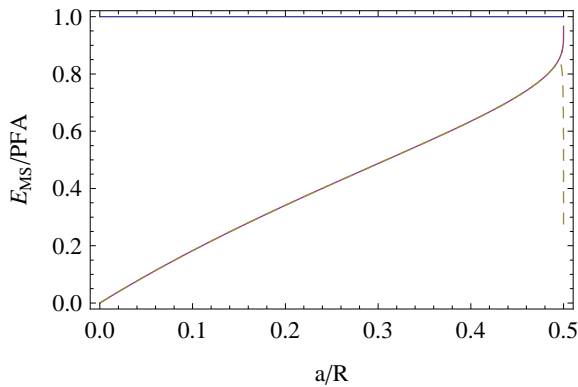}
\caption{\label{fig3}
Plotted is the ratio of the exact interaction energy
(\ref{exsphere}) of two weakly-coupled
spheres to the proximity force approximation
(\ref{pfawcsphere}) as a function of the sphere radius $a$
for $a=b$.  Shown also by a dashed line is the power series expansion
(\ref{mesphere}), truncated at
$n=100$, indicating that it is necessary to include
very high powers to capture the proximity force limit.}
\end{figure}

\begin{figure}[t]
\vspace{2in}
\includegraphics{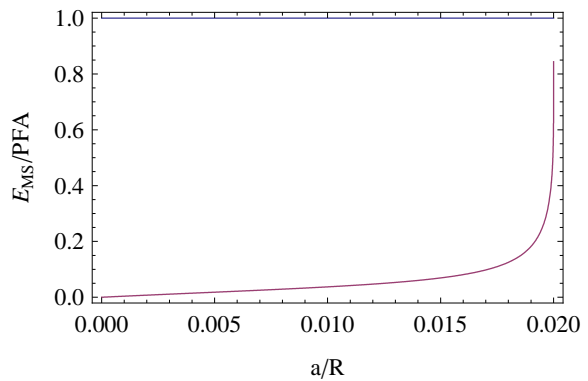}
\caption{\label{fig4}
Plotted is the ratio of the exact interaction energy
(\ref{exsphere}) of two weakly-coupled
spheres to the proximity force approximation
(\ref{pfawcsphere}) as a function of the sphere  radius $a$
for $b/a=49$.}
\end{figure}

\section{Conclusion}
We have used standard multiple scattering techniques to calculate
the Casimir interaction between two semitransparent ($\delta$-function)
spheres and between two semitransparent parallel cylinders.  When the
coupling constant is weak, we are able to sum the power series expansion
in $a/R$ 
exactly, and obtain a closed form for the Casimir interaction energy.
This energy reduces to the proximity force limit when the bodies are
very close together.  Formulas which extrapolate away from the limiting
proximity force theorem are commonly referred to as proximity force
approximations; these are ambiguous and
in general, the PFAs do a poor job in describing
the interaction.  These exact results represent the first known exact
closed-form results for the Casimir 
interaction between two bodies which are not plane surfaces.
More details and examples will be given in Ref.~\cite{scat}.
\begin{acknowledgments}
We thank the US National Science Foundation (Grant No.\ PHY-0554926) and the
US Department of Energy (Grant No.\ DE-FG02-04ER41305) for partially funding
this research.  We thank Prachi Parashar and K. V. Shajesh for extensive
collaborative assistance throughout this project.
  We are grateful to many participants in the workshop
on Quantum Field Theory Under the Influence of External Conditions held
in Leipzig in September 2007 (QFEXT07) for many illuminating lectures and
discussions.
\end{acknowledgments}


\begin{thebibliography}{99}

   
\bibitem{Balian:2004jv}
  R.~Balian and B.~Duplantier,
  %``Geometry of the Casimir Effect,''
  arXiv:quant-ph/0408124, 
  in the proceedings of 15th SIGRAV Conference on General Relativity and 
  Gravitational Physics, Rome, Italy, 9--12 September 2002. 
  %%CITATION = QUANT-PH/0408124;%%
  

\bibitem{Balian:1977qr}
  R.~Balian and B.~Duplantier,
  %``Electromagnetic Waves Near Perfect Conductors. 2. Casimir Effect,''
  Ann.\ Phys.\ (N.Y.)  {\bf 112}, 165 (1978).
  %%CITATION = APNYA,112,165;%%
  
 
\bibitem{Balian:1976za}
  R.~Balian and B.~Duplantier,
  %``Electromagnetic Waves Near Perfect Conductors. 1. Multiple Scattering
  %Expansions. Distribution Of Modes,''
  Ann.\ Phys.\ (N.Y.)  {\bf 104}, 300 (1977).
  %%CITATION = APNYA,104,300;%%

\bibitem{buscher} R. B\"uscher and T. Emig, 
%``Geometry and Spectrum of Casimir Forces,''
Phys.\ Rev.\ Lett.\ {\bf 94}, 133901 (2005).

\bibitem{Emig:2006uh}
  T.~Emig, R.~L.~Jaffe, M.~Kardar and A.~Scardicchio,
  %``Casimir interaction between a plate and a cylinder,''
  Phys.\ Rev.\ Lett.\  {\bf 96}, 080403 (2006)
  [arXiv:cond-mat/0601055].
  %%CITATION = PRLTA,96,080403;%%


\bibitem{Emig:2002xz}
  T.~Emig,
  %``Casimir Forces: An Exact Approach for Periodically Deformed Objects,''
  Europhys.\ Lett.\  {\bf 62}, 466 (2003)
  [arXiv:cond-mat/0206585].
  %%CITATION = EULEE,62,466;%%

%\bibitem{Sommerfeld}
%A. Sommerfeld, Ann.\ Phys.\ (Leipzig) {\bf 28}, 44 (1909).




\bibitem{Bulgac:2005ku}
  A.~Bulgac, P.~Magierski and A.~Wirzba,
%``Scalar Casimir effect between Dirichlet spheres or a plate and a  sphere,''
  Phys.\ Rev.\  D {\bf 73}, 025007 (2006)
  [arXiv:hep-th/0511056].
  %%CITATION = PHRVA,D73,025007;%%

\bibitem{Wirzba:2005zn}
  A.~Wirzba, A.~Bulgac and P.~Magierski,
 %``Casimir interaction between normal or superfluid grains in the Fermi sea,''
  J.\ Phys.\ A  {\bf 39}, 6815 (2006)
  [arXiv:quant-ph/0511057].
%%CITATION = JPAGB,A39,6815;%%

\bibitem{wirzba07}
A. Wirzba, to appear in Proceedings of QFEXT07 [arXiv:0711.2395]


\bibitem{krein}
M. G. Krein, Mat. Sb. (N.S.) {\bf 33}, 597 (1953); Dokl.\ Akad.\ Nauk
SSSR {\bf 144}, 268 (1962) [Sov.\ Math.-Dokl. {\bf 3}, 707 (1962)];
M. Sh.\ Birman and M. G. Krein, Dokl.\ Akad.\ Nauk
SSSR {\bf 144}, 475 (1962) [Sov.\ Math.-Dokl. {\bf 3}, 740 (1962)].

\bibitem{Bordag:2006vc}
  M.~Bordag,
  %``The Casimir effect for a sphere and a cylinder in front of plane and
  %corrections to the proximity force theorem,''
  Phys.\ Rev.\  D {\bf 73}, 125018 (2006)
  [arXiv:hep-th/0602295].
  %%CITATION = PHRVA,D73,125018;%%

\bibitem{Bordag:2006kx}
  M.~Bordag,
  %``Generalized Lifshitz formula for a cylindrical plasma sheet in front of a
  %plane beyond proximity force approximation,''
  Phys.\ Rev.\  D {\bf 75}, 065003 (2007)
  [arXiv:quant-ph/0611243].
  %%CITATION = PHRVA,D75,065003;%%
 
\bibitem{Mazzitelli:2006ne}
  F.~D.~Mazzitelli, D.~A.~R.~Dalvit and F.~C.~Lombardo,
  %``Exact zero-point interaction energy between cylinders,''
  New J.\ Phys.\  {\bf 8}, 240 (2006)
  [arXiv:quant-ph/0610181].
  %%CITATION = NJOPF,8,240;%%


\bibitem{Dalvit:2006wy}
  D.~A.~R.~Dalvit, F.~C.~Lombardo, F.~D.~Mazzitelli and R.~Onofrio,
  %``Exact Casimir interaction between eccentric cylinders,''
  Phys.\ Rev.\  A {\bf 74}, 020101 (2006).
  %%CITATION = PHRVA,A74,020101;%%

\bibitem{Emig:2007cf}
  T.~Emig, N.~Graham, R.~L.~Jaffe and M.~Kardar,
  %``Casimir forces between arbitrary compact objects,''
  arXiv:0707.1862 [cond-mat.stat-mech], Phys.\ Rev.\ Lett.\
{\bf99}, 170403 (2007)
  %%CITATION = ARXIV:0707.1862;%%
  

\bibitem{Emig:2007me}
  T.~Emig, N.~Graham, R.~L.~Jaffe and M.~Kardar,
  %``Casimir Forces between Compact Objects: I. The Scalar Case,''
  arXiv:0710.3084 [cond-mat.stat-mech].
  %%CITATION = ARXIV:0710.3084;%%

\bibitem{Emig:2007qw}
  T.~Emig and R.~L.~Jaffe,
  %``Casimir forces between arbitrary compact objects: Scalar and
  %electromagnetic field,''
to appear in Proceedings of QFEXT07,
  arXiv:0710.5104 [quant-ph].
  %%CITATION = ARXIV:0710.5104;%%

\bibitem{sidewalls}
S. J. Rahi, A. W. Rodriguez, T. Emig, R. L. Jaffe, S. G. Johnson,
and M. Kardar,  arXiv:0711.1987
%    Title: Nonmonotonic effects of parallel sidewalls on Casimir forces
%between cylinders

\bibitem{Kenneth:2007jk}
  O.~Kenneth and I.~Klich,
  %``Casimir forces in a T operator approach,''
  arXiv:0707.4017 [quant-ph].
  %%CITATION = ARXIV:0707.4017;%%
  
  \bibitem{lippmann}
  B. A. Lippmann and J. Schwinger, Phys.\ Rev.\ {\bf 79}, 469 (1950).

\bibitem{Gies:2006bt}
  H.~Gies and K.~Klingm\"uller,
  %``Casimir effect for curved geometries: PFA validity limits,''
  Phys.\ Rev.\ Lett.\  {\bf 96}, 220401 (2006)
  [arXiv:quant-ph/0601094].
  %%CITATION = PRLTA,96,220401;%%


\bibitem{Gies:2006xe}
  H.~Gies and K.~Klingm\"uller,
  %``Casimir edge effects,''
  Phys.\ Rev.\ Lett.\  {\bf 97}, 220405 (2006)
  [arXiv:quant-ph/0606235].
  %%CITATION = PRLTA,97,220405;%%


\bibitem{Gies:2006cq}
  H.~Gies and K.~Klingm\"uller,
  %``Worldline algorithms for Casimir configurations,''
  Phys.\ Rev.\  D {\bf 74}, 045002 (2006)
  [arXiv:quant-ph/0605141].
  %%CITATION = PHRVA,D74,045002;%%



\bibitem{Gies:2003cv}
  H.~Gies, K.~Langfeld and L.~Moyaerts,
  %``Casimir effect on the worldline,''
  JHEP {\bf 0306}, 018 (2003)
  [arXiv:hep-th/0303264].
  %%CITATION = JHEPA,0306,018;%%





\bibitem{capasso}
A. Rodrigues, M. Ibanescu, D. Iannuzzi, F. Capasso, J. D. Joannopoulos,
and S. G. Johnson, %``Computation and visualization of Casimir forces in
%arbitrary geometries: non-monotonic lateral forces and failure proximity
%force approximations,'' 
arXiv:0704.1890v2.


\bibitem{Schwinger:1977pa}
  J.~Schwinger, L.~L.~DeRaad, Jr., and K.~A.~Milton,
  %``Casimir effect in dielectrics,''
  Ann.\ Phys.\ (N.Y.)  {\bf 115}, 1 (1979).
  %%CITATION = APNYA,115,1;%%

%\cite{Milton:1978sf}
\bibitem{Milton:1978sf}
  K.~A.~Milton, L.~L.~DeRaad, Jr., and J.~Schwinger,
  %``Casimir Selfstress On A Perfectly Conducting Spherical Shell,''
  Ann.\ Phys.\ (N.Y.)  {\bf 115}, 388 (1978).
  %%CITATION = APNYA,115,388;%%

 
  \bibitem{Schwinger75}
  J. Schwinger, Lett.\ Math.\ Phys.\ {\bf 1}, 43 (1975).
  
%\bibitem{Milton:2007ar}
%  K.~A.~Milton, P.~Parashar, K.~V.~Shajesh and J.~Wagner,
%  %``How does Casimir energy fall? II. Gravitational acceleration of quantum
%  %vacuum energy,''
%J. Phys.\ A: Math.\ Theor.\ {\bf 40}, 10935 (2007),
%  arXiv:0705.2611 [hep-th].
%  %%CITATION = ARXIV:0705.2611;%%

%\bibitem{Boyer:1968uf}
%  T.~H.~Boyer,
%  %``Quantum electromagnetic zero point energy of a conducting spherical shell
%  %and the Casimir model for a charged particle,''
%  Phys.\ Rev.\  {\bf 174}, 1764 (1968).
%  %%CITATION = PHRVA,174,1764;%%


%\bibitem{barton04}
% G. Barton, J. Phys.\ A: Math.\ Gen.\ {\bf 37}, 1011 (2004).
 
  
%\bibitem{Scandurra:1998xa}
%  M.~Scandurra,
%  %``The ground state energy of a massive scalar field in the background of  a
%  %semi-transparent spherical shell,''
%  J.\ Phys.\ A: Math.\ Gen.\ {\bf 32}, 5679 (1999)
%  [arXiv:hep-th/9811164].
%  %%CITATION = JPAGB,A32,5679;%%

%\bibitem{Milton:2004vy}
%  K.~A.~Milton,
%  %``Casimir energies and pressures for delta-function potentials,''
%  J.\ Phys.\ A: Math.\ Gen.\  {\bf 37}, 6391 (2004)
%  [arXiv:hep-th/0401090].
%  %%CITATION = JPAGB,A37,6391;%%


%\cite{Milton:2004ya}
%\bibitem{Milton:2004ya}
%  K.~A.~Milton,
%  %``The Casimir effect: Recent controversies and progress,''
%  J.\ Phys.\ A: Math.\ Gen.\  {\bf 37}, R209 (2004)
%  [arXiv:hep-th/0406024].
%  %%CITATION = JPAGB,A37,R209;%%


%\bibitem{ponzano}
%G. Ponzano and T. Regge, ``Semiclassical Limit of Racah Coefficients,''
%in {\it Spectroscopic and Group Theoretical Methods in Physics} (Racah
%Memorial Volume), ed.~F. Bloch, S. G. Cohen, A De-Shalit, S. Sambursky, and
%I. Talmi (Wiley, New York, 1968), p.~1.

%\bibitem{biedenharn}
%L. C. Biedenharn and J. D. Louck, {\it The Racah-Wigner Algebra in Quantum
%Theory} (Encyclopedia of Mathematics and Its Applications, Vol.~9)
%(Addison-Wesley, Reading, MA, 1981), pp.~371--6.

%\bibitem{aquilanti}
%V. Aquilanti, H. M. Haggard, R. G. Littlejohn, and L. Yu, J. Phys.~A:
%Math.\ Theor.\ {\bf 40}, 5637 (2007).

\bibitem{scat}
K. A. Milton and J. Wagner, ``Multiple Scattering Methods in Casimir
Calculations,'' arXiv:0712.3811.


\bibitem{riordan} J. Riordan, {\em An Introduction to Combinatorial
Analysis} (Dover, Mineola, NY, 2002), p.~191.




\end{thebibliography}
\end{document}